\documentclass[11pt,twoside]{article}

\usepackage{epsfig,amsfonts,color}
\usepackage{amsmath}

\usepackage{amssymb, palatino, geometry,url}
\usepackage{algorithmic}
\usepackage[noresetcount,lined,boxed]{algorithm2e} % ... for algorithms
\newcommand{\mynote}[1]{{ \color{black}{#1}}}
\usepackage[colorlinks=true,linkcolor=blue,citecolor=blue,urlcolor=blue]{hyperref}

\usepackage{fancyhdr}
\newcommand{\be}{\begin{equation}}
\newcommand{\ee}{\end{equation}}
\newcommand{\bea}{\begin{eqnarray}}
\newcommand{\eea}{\end{eqnarray}}

\newcommand{\bvec}{\left(\begin{array}{c}}
\newcommand{\evec}{\end{array}\right)}
\newcommand{\bsub}{\begin{subequations}}
\newcommand{\esub}{\end{subequations}}

\usepackage{lineno}
\begin{document}

\title{Image-Based Model Predictive Control \\ via Dynamic Mode Decomposition}

\author{Qiugang Lu${}^{\ddag}$ and Victor M. Zavala${}^{\P}$\thanks{Corresponding Author: victor.zavala@wisc.edu}\\
  {\small ${}^{\P}$Department of Chemical and Biological Engineering}\\
 {\small \;University of Wisconsin - Madison, 1415 Engineering Dr, Madison, WI 53706, USA}}
 \date{}
\maketitle

\begin{abstract}
We present a data-driven model predictive control (MPC) framework for systems with high state-space  dimensionality. This work is motivated by the need to exploit sensor data that appears in the form of images  (e.g., 2D or 3D spatial fields captured by thermal cameras).  We propose to use dynamic mode decomposition (DMD) to directly build a low-dimensional model from image data and we use such model to obtain a tractable MPC controller. We demonstrate the scalability of this approach (which we call DMD-MPC) by using a 2D thermal diffusion system. Here, we assume that the evolution of the thermal field is captured by 50x50 pixel images, which results in a 2500-dimensional state-space. We show that that the dynamics of this high-dimensional space can be accurately captured by using a 40-dimensional DMD model and we show that the field can be manipulated satisfactorily by using an MPC controller that embeds the low-dimensional DMD model.  We also show that the DMD-MPC controller outperforms a standard MPC controller that uses data collected from a finite set of spatial locations (proxy locations) to manipulate the high-dimensional thermal field. This illustrates the value of information embedded in image data. 
\end{abstract}


\section{Introduction}

Emerging camera-based sensing and computer vision technologies generate data in the form of images and videos (sequences of images).  These capabilities have been widely used for the control of robotic systems \cite{das2002vision}, vehicle systems \cite{ghoreyshi2014reduced}, and crystallization systems \cite{larsen2006industrial} but have seen limited use in other application domains such as process control.  Exploiting these new data sources is not straightforward because such sources tend to be high-dimensional and distributed in nature (e.g., contain 2D and 3D spatial fields or spatial trajectories) and because these data sources are not directly compatible with traditional modeling and control techniques.  Specifically, most control architectures use models that are built using sensor data that is collected at a finite set of fixed spatial locations. Take the example of temperature control of a room; here, the thermostat collects data at a single location and this location is used as a {\em proxy point} to manipulate the entire temperature field of the room (an infinite-dimensional 3D space) \cite{kontes2017using}. Another example is that of the control of a distillation (separation) tower; here, temperature sensors at selected locations are used as proxies to manipulate the internal temperature field (an infinite-dimensional 3D space) \cite{gilles1980reduced}. In fact, one could argue that, {\em all systems are inherently infinite-dimensional} (they live in continuous 3D spaces) but we treat them as low-dimensional systems in order to enable control with existing sensing and actuation technologies. 

Now imagine that we had the ability to observe the entire 3D thermal field of a room or of a column by using a thermal camera. How could one exploit this information to manipulate the field?  Here, one could envision deriving a physical model that has a state-space representation that is compatible with that of the spatial image data. Unfortunately, such models tend to be complex (e.g., involve partial differential equations \cite{baker2000finite,qi2014multi}) and can be difficult to handle computationally by using control techniques such as MPC \cite{zavala2009optimization}. To deal with this issue, one could use a model reduction technique such as proper orthogonal decomposition or balanced truncation  \cite{moore1981principal,juang1985eigensystem,rowley2005model,antoulas2005approximation} to construct a low-dimensional representation of the physical model. We could then embed such a low-order model within a tractable MPC formulation \cite{narasingam2017development}.  However, building detailed physics-based models is time-consuming or, even worse, capturing the level of detail that image data captures (e.g., complex geometries) might simply not be practical to do. To overcome this obstacle, one could envision using the image data to directly build an empirical (data-driven) dynamical model \cite{favoreel2000subspace}. Unfortunately, it might not be possible to build a model of the same dimension as that of the spatial field captured by the image (which is inherently high-dimensional). This is because data-driven models tend to be dense and it would be difficult to store them and to perform computations with them (e.g., compute MPC control action) \cite{zavala2008fast,frison2016efficient}.    

\begin{figure}[!htp]
	\begin{center}
		\includegraphics[width=0.9\textwidth]{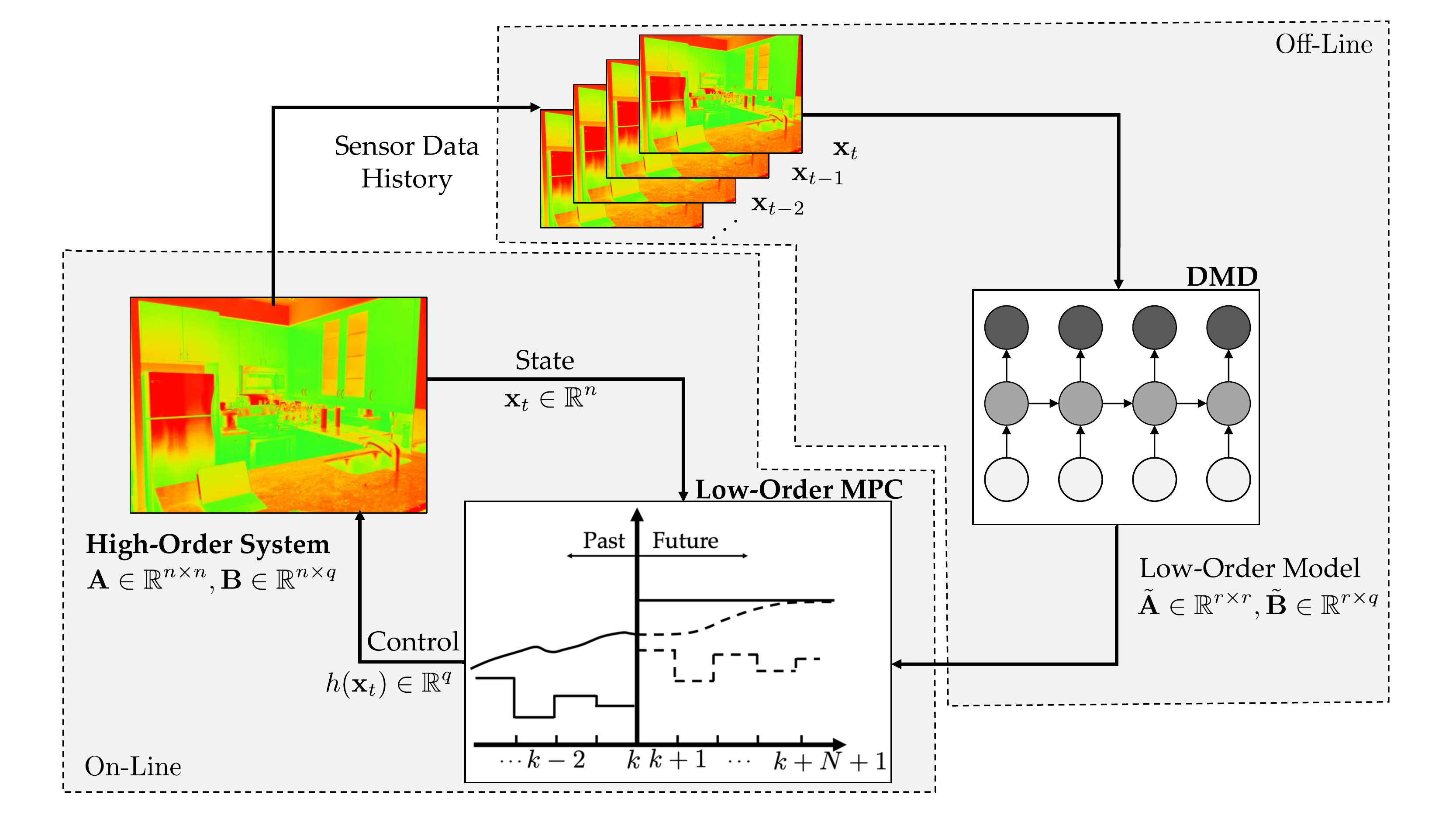}\caption{Data-driven MPC framework based on low-dimensional DMD models (DMD-MPC).}\label{fig: Illustration}
	\end{center}
\end{figure}

In this work, we present a {\em data-driven MPC} framework (which we call DMD-MPC). The framework uses low-dimensional models that are directly built from image data (see Figure \ref{fig: Illustration}). Here, we focus our attention on a modeling technique that is known as dynamic mode decomposition (DMD) \cite{kutz2016dynamic}. DMD exploits the fact that high-dimensional dynamical systems tend to reside in low-dimensional spaces and thus seeks to construct reduced order models in such spaces \cite{schmid2010dynamic}.  In the context of spatio-temporal systems (e.g., diffusion and fluid flow), the low-dimensional space is composed of modes that capture dominant spatial features of the system. DMD modes can also be interpreted as approximations of Koopman operator modes for nonlinear systems \cite{mezic2013analysis}. Notably, DMD can extract such dominant modes directly from data (i.e., from snapshots of the spatial field) to build low-dimensional models. This is done by finding a low-rank dynamic mapping that minimizes the prediction error, which can be obtained by solving a rank-constrained regression problem \cite{narasingam2017development,noack2008finite}. As a result, a DMD model has optimality guarantees (in that this is the low-rank model that best predicts the data). We have recently shown that the approximation error of DMD models has desirable asymptotic properties; specifically, we have shown that the prediction error vanishes as one increases data availability and model order \cite{lu2020characterizing}.  Moreover, we have recently established that DMD is equivalent to subspace identification (most prevalent data-driven modeling technique) \cite{shin2020unifying}.  

DMD has seen many  applications in areas such as video processing \cite{kutz2016dynamicVideo}, fluid dynamics \cite{tissot2014model}, and financial time series \cite{mann2016dynamic}. Many variants of DMD have been proposed to improve the robustness and application scope of DMD  \cite{kou2019dynamic}. Examples include optimized DMD \cite{chen2012variants,wynn2013optimal,shin2020unifying}, sparsity-promoting DMD \cite{jovanovic2014sparsity} (leads to sparse models), and extended DMD \cite{williams2015data} (can handle high-order dynamics). A number of methods for selecting modes have also been proposed \cite{bistrian2015improved,kou2017improved}. DMD models can also be constructed to handle non-autonomous systems \cite{proctor2016dynamic}. In this context, a key benefit of DMD is that it delivers a low-dimensional (tractable) model that can be embedded in MPC formulations. Recent work has proposed to use DMD models in MPC to perform wind farm control \cite{annoni2016wind} and hydraulic fracturing control \cite{narasingam2017development}. We note that building low-dimensional  models from data can also be accomplished by using powerful emerging techniques such as autoencoders \cite{lee2020model}. These approaches can handle more complex representations than DMD (e.g., nonlinear) but are more difficult to handle computationally in MPC formulations. Moreover, their theoretical properties (e.g., behavior of approximation errors) are still not well-understood. 

The availability of image data, together with emerging modeling techniques such as DMD, opens new opportunities to design powerful MPC architectures.  We demonstrate these capabilities by applying our DMD-MPC architecture to  a 2D thermal diffusion system. Here, we show that DMD can extract the dominant spatial modes of the physical system from snapshots of the thermal field.  We also show that a DMD model of dimension 40 gives accurate predictions of a system model that has a dimension of 2500 (reduction of 98\%). The DMD model enables a tractable MPC formulation and we show that this formulation provides satisfactory tracking and constraint satisfaction performance. We show that the DMD-MPC controller drastically outperforms a standard MPC controller that uses data from a finite set of (proxy) spatial locations to manipulate the thermal field. We attribute these results to the fact that DMD modes more effectively capture information of the entire field (compared to data at fixed locations).  These results thus illustrate the value of information embedded in image data.  Our work also seeks to point out  a number of challenges that need to be addressed to effectively use image data in MPC. These challenges include dealing with inherent limitations in actuation and controllability, dealing with nonlinear systems, and dealing with mixed data-driven and physics-driven models. 

The paper is structured as follows: In Section \ref{sec:DMDc}, we introduce concepts and computational aspects of DMD models. Section \ref{sec: MPCDMDc} presents the proposed MPC framework based on DMD models. Section \ref{sec: DMDDiffusion} provides a 2D thermal diffusion system to illustrate the developments. The paper closes in Section \ref{sec: Summary} with concluding remarks and a perspective on  open challenges.

%%%%%%%%%%%%%%%%%%%%%%%%%%%%%%%%%%%%%%%%%%
\section{Dynamic Mode Decomposition}\label{sec:DMDc}

Consider a sequence of $m$ time snapshots of the state vector $\mathbf{x}\in \mathbb{R}^{n}$.  We represent the state sequence as $\{\mathbf{x}_{1},\mathbf{x}_{2},\ldots,\mathbf{x}_{m}\}$.  Associated with each state snapshot, there is a measurement of the control vector $\mathbf{u}\in\mathbb{R}^{q}$ and we denote the control sequence as $\{\mathbf{u}_{1},\mathbf{u}_{2},\ldots,\mathbf{u}_{m}\}$. The image-based setting that we consider here involves $n\gg 1$ (high-dimensional state space) and $q\ll n$ (much fewer actuators than observed states). Specifically, an image captures a spatial field that might not be possible to fully manipulate using available actuators. This issue is common in the context of control of systems described by partial differential equations (e.g., actuators only exist at the domain boundaries) \cite{baker2000finite}. 

An image is a finite-dimensional (discrete) representation of an infinite-dimensional (continuous) field.  We represent an image as the matrix $\mathbb{X}\in \mathbb{R}^{p\times p}$. Each row-column position is a pixel and the entry at such position is known as the intensity.  We note that the image only represents a scalar field (it represents a single state variable). For instance, the image might capture the spatial distribution of temperature (the row-column position is the spatial location and the intensity is the temperature level).  The state is the vectorized form of the image and we thus have that $\mathbf{x}=\textrm{vec}(\mathbb{X})$ with $n=p^2$. The dimension $n$ is thus given by the total number of pixels in the image. An image can also be a superposition of color channels (e.g., RGB or L$*$a$*$b) in which each channel has an associated matrix with $p^2$ pixels. In such a case, the state vector $\mathbf{x}$ is a vector containing all the color channel matrices and has a dimension $n= 3\cdot p^2$.    Similarly, it is possible for the state vector $\mathbf{x}$ to capture multiple spatial fields (e.g., temperature and flow). 

We assume that the true system is described by a linear model: 
\begin{equation}
\mathbf{x}_{k+1}=\mathbf{A}\mathbf{x}_{k} + \mathbf{B} \mathbf{u}_k, \label{eq: LinearModel}
\end{equation}
where $\mathbf{A}\in\mathbb{R}^{n\times n}$ is the system matrix, and its eigenvalues characterize the dynamical behavior of the state. The input matrix $\mathbf{B}\in\mathbb{R}^{n\times q}$ captures the effects of the control inputs on the system state. In our setting, both $\mathbf{A}$ and $\mathbf{B}$ are unknown (need to be estimated from data) and high-dimensional (compatible with the dimension of the image). 

We split the data snapshots as:
\begin{align}
\mathbf{X}&:=[\mathbf{x}_{1},\mathbf{x}_{2},\ldots,\mathbf{x}_{m-1}],\quad \mathbf{Y}:=[\mathbf{x}_{2},\mathbf{x}_{3},\ldots,\mathbf{x}_{m}], \quad 
\mathbf{\Upsilon} := [\mathbf{u}_{1},\mathbf{u}_{2},\ldots,\mathbf{u}_{m-1}],
\end{align}
and we thus have that:
\begin{align}
\mathbf{Y}=\mathbf{A}\mathbf{X}+\mathbf{B}\mathbf{\Upsilon}=\mathbf{\Theta}\mathbf{\Omega},
\end{align}
where $\mathbf{\Theta}:=[\mathbf{A}~\mathbf{B}]$ are the model parameters and $\mathbf{\Omega}:=[\mathbf{X}^{T}~\mathbf{\Upsilon}^{T}]^{T}$ is the data matrix. In DMD, the model $\mathbf{\Theta}$ is estimated by minimizing the residual $\|\mathbf{Y}-\mathbf{\Theta \Omega}\|_2^2$ (prediction error). The solution is:
\begin{equation}
\hat{\mathbf{\Theta}}=\mathbf{Y}\mathbf{\Omega}^{\dagger},
\end{equation}	
where ${\dagger}$ is the Moore-Penrose pseudoinverse. The pseudoinverse can be computed using SVD \cite{schmid2010dynamic};  here, the data matrix is decomposed as:
\begin{equation}
\mathbf{\Omega}\approx\hat{\mathbf{U}}_s\hat{\mathbf{\Sigma}}_s \hat{\mathbf{V}}_s^{T}, \label{eq: SVD_Omega}
\end{equation}	
where $s$ is the truncation order in the economy-size SVD, $\hat{\mathbf{U}}_s\in\mathbb{R}^{(n+q)\times s}$, $\hat{\mathbf{\Sigma}}_s\in\mathbb{R}^{s\times s}$, and $\hat{\mathbf{V}}_s\in\mathbb{R}^{(m-1)\times s}$. Spatial fields often contain a few coherent structures and as a result, $s \ll n$. In practice, $s$ is determined by a prescribed threshold and the modes with singular values greater than the threshold will be reserved. It is also observed that the columns in $\hat{\mathbf{U}}_s$ are not the first $s$ principal bases from the data matrix $\mathbf{X}$. Instead, they are the first $s$ modes of the joint data matrix between $\mathbf{X}$ and $\mathbf{\Upsilon}$. The estimated system and input matrices can thus be expressed as: 
\begin{align}
\hat{\mathbf{A}}=\mathbf{Y}\hat{\mathbf{V}}_{s}\hat{\mathbf{\Sigma}}_{s}^{-1}\mathbf{\hat{U}}_{1,s}^{T}, ~ \hat{\mathbf{B}}=\mathbf{Y}\hat{\mathbf{V}}_{s}\hat{\mathbf{\Sigma}}_{s}^{-1}\hat{\mathbf{U}}_{2,s}^{T},  \label{eq: FullOrderEstimates}
\end{align}
with $\hat{\mathbf{U}}_{1,s}\in \mathbb{R}^{n\times s}$, $\hat{\mathbf{U}}_{2,s}\in \mathbb{R}^{q\times s}$, and  $\hat{\mathbf{U}}^{T}_{s}=[\hat{\mathbf{U}}^{T}_{1,s}~\hat{\mathbf{U}}^{T}_{2,s}]$. The estimated  state-space model is:
\begin{equation}
\mathbf{x}_{k+1} = \hat{\mathbf{A}} \mathbf{x}_{k} + 
\hat{\mathbf{B}} \mathbf{u}_{k}.  \label{eq: EstimatedModel}
\end{equation}
We note that $\hat{\mathbf{A}}$ is a large-scale but rank-deficient  matrix. The eigenvalues of $\hat{\mathbf{A}}$ can yield dominant dynamic modes of the system, and the computation of the eigen-decomposition of $\hat{\mathbf{A}}$  is prohibitive due to the high dimensions. Therefore, reduced models are used to approximate the original system \eqref{eq: LinearModel}, which is achieved by projecting the state $\mathbf{x}_k$ onto a low-dimensional subspace. Different from DMD for autonomous systems, the left singular vectors in $\hat{\mathbf{U}}_s$ are not appropriate to be used as the model reduction basis. Instead, one is interested in the principal component space of the state observations, and uses it as the linear subspace to project the state. To this end, the bases for the state observations $\mathbf{Y}$ are utilized to find the linear transformation and its truncated SVD is shown as \cite{kutz2016dynamic}:
\begin{equation}
\mathbf{Y}\approx\mathbf{U}_{r}\mathbf{\Sigma}_{r}\mathbf{V}_{r}^{T}, \label{eq: SVD_Y}
\end{equation}
where $r\ll n$ is the truncation order. It is likely that the truncation order $r<s$, since the  SVD on $\mathbf{Y}$ is applied with respect to the state snapshots while the former SVD \eqref{eq: SVD_Omega} accounts for the variations in both state and input. In \eqref{eq: SVD_Y}, $\mathbf{U}_r\in\mathbb{R}^{n\times r}$ provides dominant spatial modes and $\mathbf{Y}\in\mathbb{R}^{(m-1)\times r}$ provides dominant temporal patterns in data matrix $\mathbf{Y}$. Therefore, we use the columns in $\mathbf{U}_r$ (which is a non-square semi-orthogonal matrix) as the bases for the linear transformation of the true state $\mathbf{x}_{k}$.

We define the state in the low-dimensional subspace as $\tilde{\mathbf{x}}_k \in \mathbb{R}^{r}$ and the reconstructed state in the original space as $\hat{\mathbf{x}}_{k}= \mathbf{U}_r \tilde{\mathbf{x}}_{k}$. We also define the approximation error as $\mathbf{e}_k=\hat{\mathbf{x}}_{k}-{\mathbf{x}}_{k}$. The low-order approximation of model \eqref{eq: EstimatedModel} has the form: 
\begin{equation}
\tilde{\mathbf{x}}_{k+1} = \tilde{\mathbf{A}} \tilde{\mathbf{x}}_{k} + \tilde{\mathbf{B}} \mathbf{u}_{k}, \label{eq: ROM}
\end{equation}
where $\tilde{\mathbf{x}}_{k}\in \mathbb{R}^{r}$, $\tilde{\mathbf{A}}\in\mathbb{R}^{r\times r}$, $\tilde{\mathbf{B}}\in\mathbb{R}^{r\times q}$, with expressions:
\begin{subequations}
\begin{align}
\tilde{\mathbf{A}} &= \mathbf{U}_{r}^{T}\hat{\mathbf{A}}\mathbf{U}_{r}=\mathbf{U}_{r}^{T}\mathbf{Y}\hat{\mathbf{V}}_{s}\hat{\mathbf{\Sigma}}_{s}^{-1}\mathbf{\hat{U}}_{1,s}^{T}\mathbf{U}_{r}, \label{eq: ReducedOrderEstimateAtilde}\\  \tilde{\mathbf{B}} &= \mathbf{U}_{r}^{T}\hat{\mathbf{B}}=\mathbf{U}_{r}^{T}\mathbf{Y}\hat{\mathbf{V}}_{s}\hat{\mathbf{\Sigma}}_{s}^{-1}\hat{\mathbf{U}}_{2,s}^{T}. \label{eq: ReducedOrderEstimateBtilde}
\end{align}
\end{subequations}
The major eigenvalue and eigenvector pairs $(\mathbf{\Lambda},\mathbf{\Phi})$ of the system matrix $\mathbf{A}$ are assumed to be approximated sufficiently by $\hat{\mathbf{A}}$ and are computed through the eigendecomposition of $\tilde{\mathbf{A}}$ as:
\begin{equation}
\tilde{\mathbf{A}}=\mathbf{W\Lambda W}^{-1}, \quad \mathbf{\Phi}=\mathbf{U}_{r}\mathbf{W}. \label{eq: DMDEigenvectors}
\end{equation}
We adopt the approach above to calculate the eigenvectors $\mathbf{\Phi}$ (this is  known as projected DMD \cite{schmid2010dynamic}). The exact DMD method, proposed by \cite{tu2014dynamic}, involves a more complex but exact expression of the eigenvectors. Notice that computing the high-dimensional matrices $\hat{\mathbf{A}}$ and $\hat{\mathbf{B}}$ is not necessary for obtaining a reduced (low-dimensional) model. This is a key advantage over other model reduction methods such as balanced truncation that enables the construction of tractable MPC formulations.  

We have recently established that the previous procedure delivers a low-dimensional model that has consistent approximation properties \cite{lu2020characterizing}. Specifically, we have established that the approximation error $\lim_{k\to\infty}~\mathbf{e}_{k}$ vanishes as we increase the amount of data $m$ and as we increase the truncation orders $r$ and $s$. To ensure completeness, here we summarize these results.   A key assumption in the analysis is that the real system is asymptotically stable (in the absence of inputs). Specifically, for \eqref{eq: LinearModel}, let $\{\lambda_1,\ldots,\lambda_{n}\}$ be the eigenvalues of $\mathbf{A}$. We assume that $\mathbf{A}$ is Hurwitz with spectral radius:
\begin{equation}
\rho(\mathbf{A}):=\max_{i=1,\ldots,n} |\lambda_{i}| < 1. \label{eq: Hurwitz}
\end{equation}
Stability is needed because it will be shown that the dynamics of the error $\mathbf{e}_{k}$ depend on the eigenvalues of $\mathbf{A}$. The system matrix $\mathbf{A}$ being Hurwitz ensures that the error $\mathbf{e}_{k}$ decays asymptotically over time (in the absence of inputs). 

Consider \textcolor{blue}{the} real system \eqref{eq: LinearModel} and its low-order approximation \eqref{eq: ROM} obtained from DMD. Under our stability assumption, the prediction error $\mathbf{e}_{k}$ (for $k>m$) can be bounded as:
\begin{align}
&\|\mathbf{e}_{k}\| \le  M\bar{\rho}^{k-m} \|\mathbf{e}_{m}\|+M(k-m)\bar{\rho}^{k-1-m}(M_{s,m}+M_{r,m})\| \mathbf{x}_{m}\| \nonumber \\
&\qquad\qquad\qquad\qquad+M(\varepsilon_{s}^{B}+\varepsilon_{r}^{B})\sum\nolimits_{i=0}^{k-1-m}  \bar{\rho}^{k-1-m-i}\|\mathbf{u}_{i+m}\|  \nonumber \\ & \qquad\qquad\qquad\qquad +M (M_{s,m}+ M_{r,m})\sum\nolimits_{i=0}^{k-2-m}(i+1)\bar{\rho}^{i}\|\mathbf{B}\mathbf{u}_{k-2-i}\|, \label{eq: Thm2}
\end{align} 
where $\bar{\rho}\in(\rho(\mathbf{A}),1)$, $M>0$ is a bounding constant, and the positive constants $M_{s,m}$, $M_{r,m}$, $\varepsilon_{r}^B$, and $\varepsilon_{s}^B$ decrease as the SVD truncation orders $r$, $s$, and the sample size $m$ increase.  To see the asymptotic behavior we set $k \to \infty$ and find that:
\begin{equation}
\lim_{k\to\infty}\|\mathbf{e}_{k}\| \le \frac{M\bar{u}}{1-\bar{\rho}}(\varepsilon_{s}^{B}+\varepsilon_{r}^{B})+\frac{M\|\mathbf{B}\|\bar{u}}{(1-\bar{\rho})^2}(M_{s,m}+M_{r,m}). \label{eq: Thm2_limit}
\end{equation}
This result clearly highlights that the error can only converge to zero if the original system is stable. If stability holds, we have that the error vanishes as $r,s,m$ increase. This is because the constants 
$M_{s,m}$, $M_{r,m}$, $\varepsilon_{r}^B$, and $\varepsilon_{s}^B$ decrease as $r,s,m$ increase.  

We provide a sketch of the derivations of expressions \eqref{eq: Thm2}-\eqref{eq: Thm2_limit}, details can be found in \cite{lu2020characterizing}. Note that $\mathbf{e}_{k}=\mathbf{x}_{k}-\hat{\mathbf{x}}_k=\mathbf{x}_{k}-\mathbf{U}_r\tilde{\mathbf{x}}_{k}$;  from \eqref{eq: LinearModel} and \eqref{eq: ROM} we have,
\begin{align*}
\mathbf{e}_{k}&=\mathbf{A}(\mathbf{x}_{k-1}-\hat{\mathbf{x}}_{k-1})+\mathbf{A}\hat{\mathbf{x}}_{k-1}+\mathbf{B}\mathbf{u}_{k-1}-\mathbf{U}_{r}\tilde{\mathbf{A}}\tilde{\mathbf{x}}_{k-1}-\mathbf{U}_{r}\tilde{\mathbf{B}}\mathbf{u}_{k-1} \\
&=\mathbf{A}\mathbf{e}_{k-1}+(\mathbf{A}-\mathbf{U}_{r}\tilde{\mathbf{A}}\mathbf{U}_{r}^{T})(\mathbf{x}_{k-1}-\mathbf{e}_{k-1})+(\mathbf{B}-\mathbf{U}_{r}\tilde{\mathbf{B}})\mathbf{u}_{k-1} \\
&=\mathbf{U}_{r}\tilde{\mathbf{A}}\mathbf{U}_{r}^{T}\mathbf{e}_{k-1}+(\mathbf{A}-\mathbf{U}_{r}\tilde{\mathbf{A}}\mathbf{U}_{r}^{T})\mathbf{x}_{k-1}+(\mathbf{B}-\mathbf{U}_{r}\tilde{\mathbf{B}})\mathbf{u}_{k-1} \\
&=\mathbf{\Phi}\mathbf{\Lambda}\mathbf{\Phi}_{lf}^{-1}\mathbf{e}_{k-1}+\mathbf{\Psi}_{k},
\end{align*}
where the last equation follows from \eqref{eq: DMDEigenvectors}, with $\mathbf{\Psi}_{k}=(\mathbf{A}-\mathbf{\Phi}\mathbf{\Lambda}\mathbf{\Phi}_{lf}^{-1})\mathbf{x}_{k-1} +(\mathbf{B}-\mathbf{U}_{r}\tilde{\mathbf{B}})\mathbf{u}_{k-1}$ and  $\mathbf{\Phi}_{lf}^{-1}:=\mathbf{W}^{-1}\mathbf{U}_{r}^{T}$ being the left inverse of $\mathbf{\Phi}$. If we propagate the dynamics we obtain $
\mathbf{e}_{k} = \mathbf{\Phi \Lambda}^{k-m} \mathbf{\Phi}^{-1}_{lf} \mathbf{e}_{m} + \sum\nolimits_{i=0}^{k-1-m} \mathbf{\Phi} \mathbf{\Lambda}^{i} \mathbf{\Phi}^{-1}_{lf} \mathbf{\Psi}_{k-1-i}.$ Then we have
\begin{align}
	\|\mathbf{e}_{k}\| &\le M\bar{\rho}^{k-m} \|\mathbf{e}_{m}\| + M\sum\nolimits_{i=m}^{k-1} \bar{\rho}^{k-1-i} \|\mathbf{\Psi}_{i}\|.
	\label{eq: ek_norm}
\end{align}
Here, $\mathbf{\Psi}_i$ can be expressed as
\begin{align}
\mathbf{\Psi}_{i} &=(\mathbf{A}-\hat{\mathbf{A}})\mathbf{x}_{i}+ (\hat{\mathbf{A}}-\mathbf{\Phi} \mathbf{\Lambda} \mathbf{\Phi}^{-1}_{lf}) \mathbf{x}_{i}+(\mathbf{B}-\hat{\mathbf{B}})\mathbf{u}_{i}   +(\hat{\mathbf{B}}-\mathbf{U}_{r}\tilde{\mathbf{B}})\mathbf{u}_{i} \nonumber\\
&=(\mathbf{A}-\hat{\mathbf{A}})\mathbf{A}^{i-m}\mathbf{x}_{m}+\sum\nolimits_{j=0}^{i-1-m}(\mathbf{A}-\hat{\mathbf{A}})\mathbf{A}^{j}\mathbf{B}\mathbf{u}_{i-1-j} + (\hat{\mathbf{A}}-\mathbf{\Phi} \mathbf{\Lambda} \mathbf{\Phi}^{-1}_{lf}) \mathbf{A}^{i-m}\mathbf{x}_{m} + \nonumber \\
& \quad   \sum\nolimits_{j=0}^{i-1-m}(\hat{\mathbf{A}}-\mathbf{\Phi} \mathbf{\Lambda} \mathbf{\Phi}^{-1}_{lf}) \mathbf{A}^{j}\mathbf{B}\mathbf{u}_{i-1-j} + (\mathbf{B}-\hat{\mathbf{B}})\mathbf{u}_{i} + (\mathbf{I}-\mathbf{U}_{r}\mathbf{U}_{r}^{T})\hat{\mathbf{B}}\mathbf{u}_{i} \label{Psi}.
\end{align}
Note that the reconstruction error $\|\mathbf{A}-\hat{\mathbf{A}}\|$ decreases over order $s$ and $\|\mathbf{B}-\hat{\mathbf{B}}\| \le \varepsilon_{s}^B$ where $ \varepsilon_{s}^B>0$ decreases over $s$. From \eqref{eq: Hurwitz} it follows that $\|(\mathbf{A}-\hat{\mathbf{A}})\mathbf{A}^{i}\| \le M_{s,m}\bar{\rho}^{i}$, $\|(\hat{\mathbf{A}}-\mathbf{\Phi} \mathbf{\Lambda} \mathbf{\Phi}^{-1}_{lf})\mathbf{A}^{i}\| \le M_{r,m} \bar{\rho}^{i}$, $\|(\mathbf{I}-\mathbf{U}_{r}\mathbf{U}_{r}^{T})\hat{\mathbf{B}}\| \le \varepsilon_{r}^{B}$, where $M_{s,m}>0$, $M_{r,m}>0$  are constants decreasing as $s$, $r$, or $m$ increases. The constant $\varepsilon_{r}^{B}>0$ decreases as $r$ increases. Combining these results yields:
\begin{align}
&\sum_{i=m}^{k-1} \bar{\rho}^{k-1-i} \|\mathbf{\Psi}_{i}\|\\
&\le  (k-m)\bar{\rho}^{k-1-m}(M_{s,m}+M_{r,m})\|\mathbf{x}_m\| \nonumber \\
& + (\varepsilon_{s}^{B}+\varepsilon_{r}^{B})  \sum_{i=0}^{k-1-m} \bar{\rho}^{k-1-m-i} \|\mathbf{u}_{i+m}\|+(M_{s,m}+M_{r,m})\cdot \sum_{i=0}^{k-1-m}\sum_{j=0}^{i-1}\bar{\rho}^{k-1-m-i+j} \|\mathbf{B}\mathbf{u}_{i+m-1-j}\|. \label{eq: Psi_norm}
\end{align}
By using a change of variables, the main result \eqref{eq: Thm2} can be obtained. To establish 
\eqref{eq: Thm2_limit}, take $k\to\infty$, then the first two terms in  \eqref{eq: Thm2} vanish. For the last two terms, using $\lim_{l\to\infty} \sum_{k=0}^{l}z^{k}=1/(1-z)$ and $\lim_{l\to\infty} \sum_{k=0}^{l}(k+1)z^{k}=1/(1-z)^2$, for all $|z|<1$, we can obtain \eqref{eq: Thm2_limit}.
\\

\textbf{Remark 1.} The prediction error analysis for the low-order model from DMD above is obtained under a linear high-dimensional system \eqref{eq: LinearModel}. However, it is noted that DMD is can be applied for approximating nonlinear dynamical systems. In fact, DMD has a close connection with the Koopman operator, which converts a finite nonlinear system into a linear infinite-dimensional system. For a nonlinear system with low-rank dynamics, the Koopman modes and eigenvalues can capture the dominant spatial and dynamical features \cite{tu2014dynamic}. The extracted DMD modes correspond to the dominant Koopman modes \cite{kutz2016dynamic}. Therefore, the developed control scheme based on the low-order models from DMD in this paper can be used to handle certain nonlinear dynamical systems.
\\

\textbf{Remark 2}. In practice, the presence of measurement noise in the data (e.g., images) can complicate the identifiability of dominant modes since small modes can be inflated by noise. As a result, there may not be a clear cutoff between the modes caused by low-rank dynamics and those caused by the noise, especially when the noise level is high. As such, it becomes difficult to determine a suitable order for the DMD model. Moreover, DMD will require more data samples to generate an accurate model due to the reduced signal-to-noise ratio caused by the measurement noise in the data. Exploring the impact of noise is an interesting topic of future work.
	
\section{DMD-MPC} \label{sec: MPCDMDc}

Because data-driven models are dense, the computational complexity of the MPC optimization problem scales  cubically with the number of states. This motivates the use of accurate low-dimensional models to formulate the MPC controller.  The aim of our MPC controller is to steer the high-dimensional state $\mathbf{x}$ to a desired target state ${\mathbf{x}}^*\in \mathbb{R}^n$. In our context, this is equivalent to seeking to steer the spatial image field to a desired field.  An issue that arises here is that we expect to only have a finite number of actuators available to manipulate the field. As a result, the system will have inherent controllability limitations.  This reveals a couple of important issues that arise when seeking to exploit image sensor data: 

\begin{itemize}
\item There is a limit to what we can actually do with image data.  Specifically, more data is expected to facilitate control but up to a certain point (and this point is limited by actuation). For instance, the fact that  we can observe the entire temperature field of a room does not imply that we can freely manipulate it because rooms often have a single actuator (an air damper). 

\item The fact that we can observe and control a finite set of points does not imply that we can control the entire field adequately.  For instance, the fact that we can control the temperature at the location of the thermostat does not imply that we can effectively manipulate the entire 3D temperature field in the room. DMD modes provide a low-dimensional representation that captures the full-dimensional field. Our hope is thus that, by controlling the low-dimensional DMD state, we can achieve better control of the entire field than that obtained with a traditional approach (in which we control the field by using a finite set of proxy locations).  However, we note that controllability obtained with DMD is limited by that of the original system. 
\end{itemize} 

The DMD basis vectors $\mathbf{U}_r$ (together with the reduced system matrix $\tilde{\mathbf{A}}$) contain the dominant modes of the dynamical system.  Given the current state $\mathbf{x}_t\in\mathbb{R}^{n}$ for the system, the MPC formulation computes the control action $h(\mathbf{x}_t)$ by solving the problem:
\begin{subequations}
\begin{align}
\min_{\mathbf{u}_{0},\ldots,\mathbf{u}_{N-1}} \quad 
&\sum_{k=0}^{N}
\|\mathbf{U}_{r}\tilde{ \mathbf{x}}_{k}-{\mathbf{x}}^*\|+\sum_{k=0}^{N-1}\|\mathbf{u}_{k}\| \label{eq: MPC_obj} \\
\textrm {s.t.} \quad &\tilde{\mathbf{x}}_{k+1} = \tilde{\mathbf{A}} \tilde{\mathbf{x}}_{k} + \tilde{\mathbf{B}} \mathbf{u}_{k}, \quad k=0,\ldots,N-1, \label{eq: MPC_model} \\
& \mathbf{u}_{k}\in \mathcal{U}, \quad k=0,\ldots,N-1, \label{eq: MPC_input} \\
& \mathbf{U}_{r}\tilde{\mathbf{x}}_{k}\in \mathcal{X}, \quad k=0,\ldots,N, \label{eq: MPC_state} \\
& \tilde{\mathbf{x}}_{0}=\mathbf{U}_{r}^{T}\mathbf{x}_{t}, \label{eq: MPC_initial}
\end{align} 
\end{subequations}
where $N$ is the length of the prediction horizon and the control action is given by $h(\mathbf{x}_t)=\mathbf{u}_0$. The dynamical model of the MPC problem is the low-dimensional system \eqref{eq: MPC_model} and we recall that we can project the reduced state $\tilde{\mathbf{x}}_k$ to the original space by using the DMD basis as $\mathbf{U}_{r}\tilde{\mathbf{x}}_k$. Moreover, we can project the current state of the system ${\mathbf{x}}_t$ to the reduced space as  $\mathbf{U}_{r}^{T}\mathbf{x}_{t}$. The set $\mathcal{U}\subseteq \mathbb{R}^{q}$ is the constraint set for the inputs and the set $\mathcal{X}\subseteq \mathbb{R}^{n}$ is the constraint set for the state (in the original high-dimensional space). Note that the state constraints are enforced indirectly through the use of the reduced model; consequently, in order to adequately handle the  constraints, it is necessary that the reduced model provides an accurate representation of the original system.  

\begin{figure}[tbh]
	\begin{center}
		\includegraphics[width=5.5in]{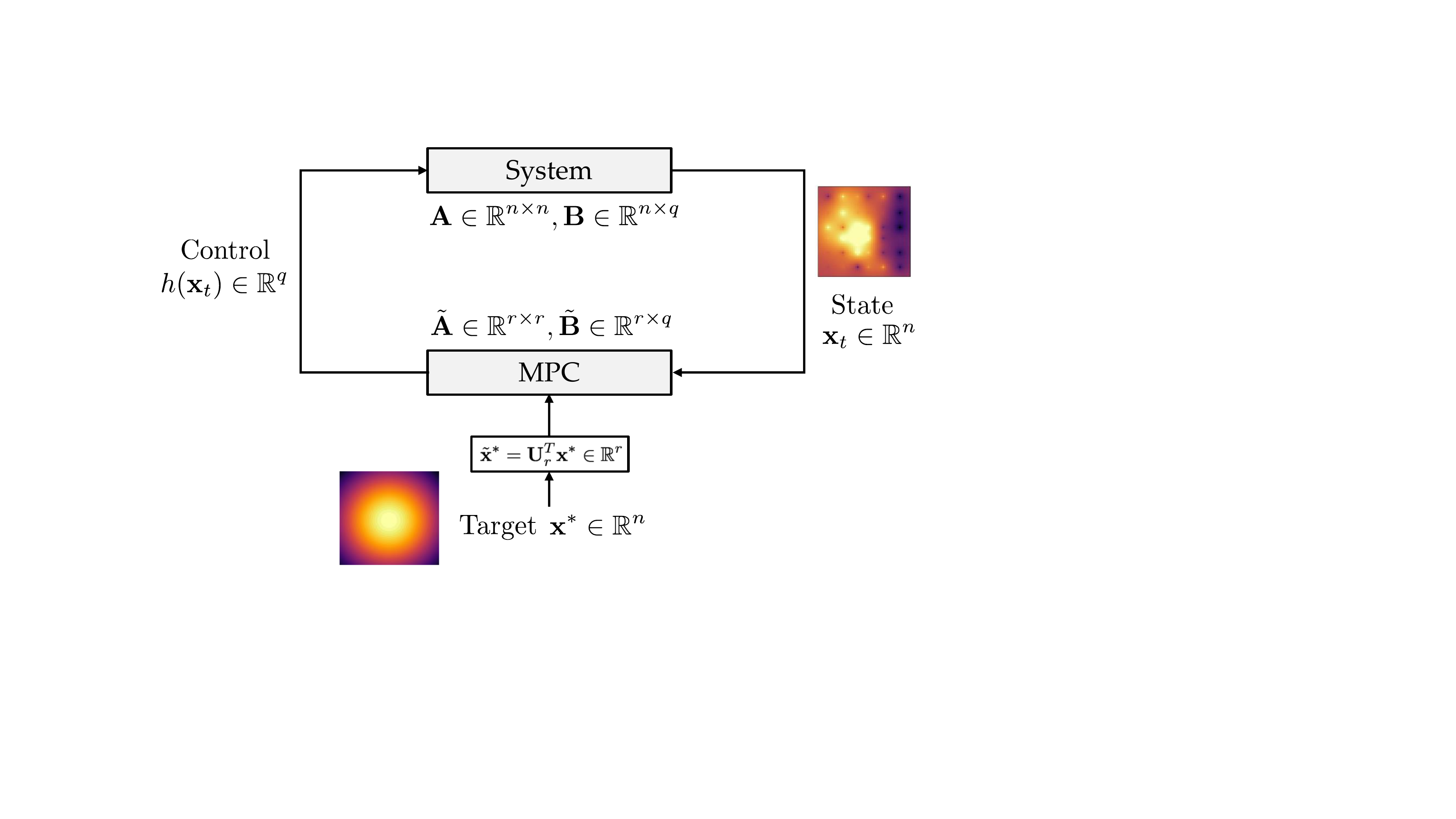}\caption{Closed-loop MPC diagram outlining dimensions of controller model and real system.}\label{fig: Diagram}
	\end{center}
\end{figure}

The closed-loop system generated by MPC is shown in Figure \ref{fig: Diagram}. Here, the MPC control action is computed using a low-dimensional (approximate) representation of the real system. The control action generates a new state for the system that is projected to the low-dimensional space of the MPC controller using the DMD basis. We note that the state tracking error in the MPC objective can also be expressed as $\|\tilde{ \mathbf{x}}_{k}-\mathbf{U}_{r}^T{\mathbf{x}}^*\|$ (because the basis $\mathbf{U}_{r}$ is orthogonal). This highlights that we can also implement the controller by projecting the target into the  low-dimensional space (without affecting the solution). We call our control architecture {\em DMD-MPC}.

\section{Case Study: 2D Thermal Diffusion System} \label{sec: DMDDiffusion}

In this section, we use a 2D diffusion system to investigate the ability of DMD-MPC to control a high-dimensional thermal field. We compare its performance to that of a standard MPC scheme that seeks to control the high-dimensional field by controlling a fixed set of spatial points (proxy locations). This comparison will help us illustrate the inherent value of image data. All scripts needed to reproduce the results can be found in \url{https://github.com/zavalab/ML/tree/master/ImageMPC}.  

\begin{figure}[!htp]
	\begin{center}
		\includegraphics[width=4in]{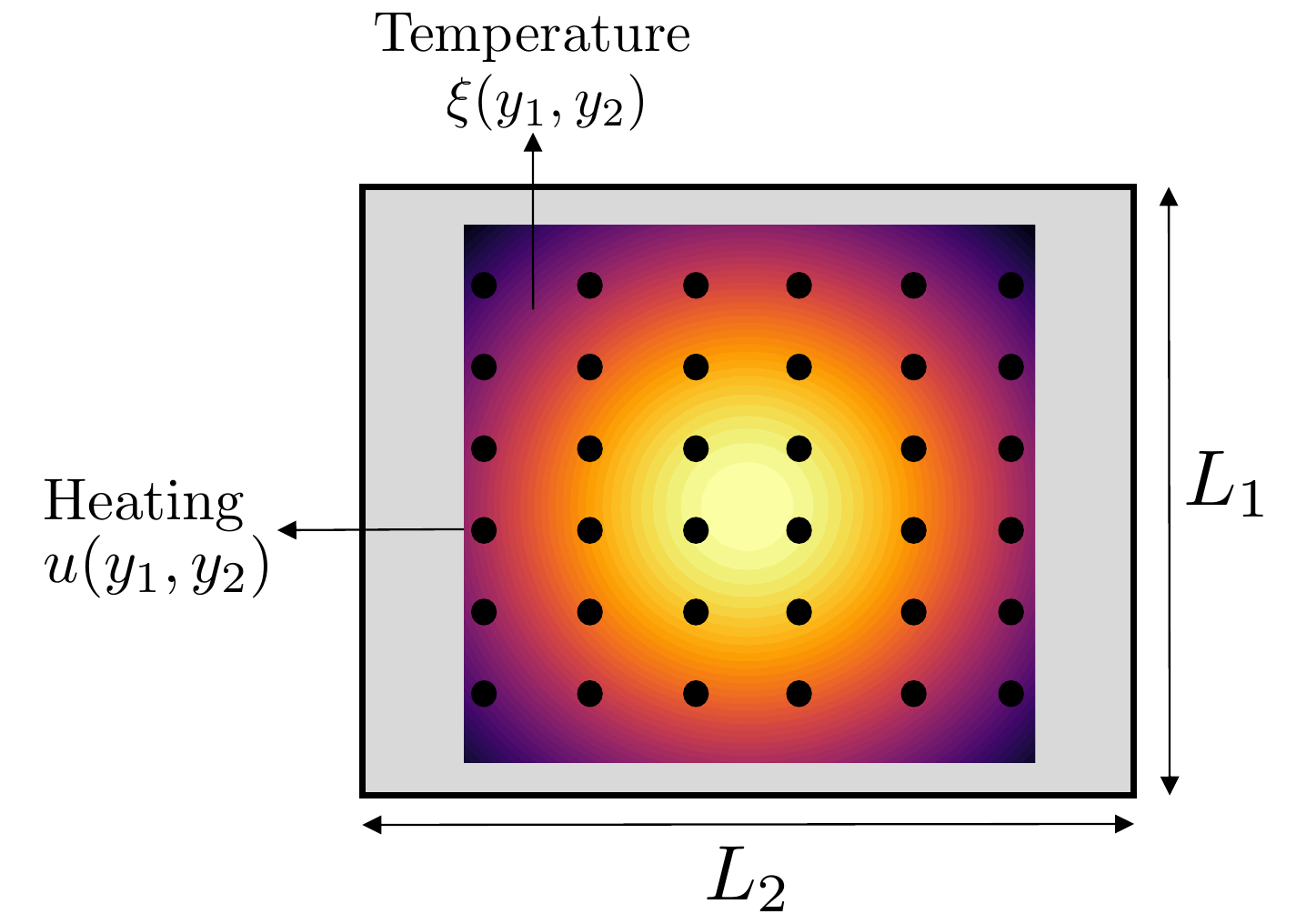}\caption{\mynote{Sketch of 2D thermal diffusion system showing complete domain (outer box) and control domain (inner box). Dots represent locations of heating devices.}}\label{fig: 2D_diffusion}
	\end{center}
\end{figure}

\subsection{Data Preparation Setup}

The system under study captures heat diffusion on a 2D field; the field is manipulated by using a set of heating devices (see Figure \ref{fig: 2D_diffusion}). This diffusion system simulates, for instance, the dynamics of a solid metal slab that is controlled by heating devices. We assume that the 2D temperature field of the system is  monitored in real time by using a camera. The goal of the DMD-MPC controller is to drive the 2D thermal field from a given state field to a target state field. The MPC controller operates with a low-dimensional representation of the field (obtained with DMD). The state of the system is reported to the controller in the form of an image. For simplicity, we assume that the image field is the field of the real system. The image field is projected down to a low-dimensional space in which the DMD-MPC controller operates. The control action computed is then fed back to the system.  

We define $\xi(y_1,y_2,t)$ as the temperature (intensity) at location $(y_1,y_2)$ and at time $t$. The thermal field is generated by using the PDE:
\begin{align}
&\frac{\partial \xi}{\partial t}=\alpha \left(\frac{\partial^2 \xi}{\partial y_1^2} + \frac{\partial^2 \xi}{\partial y_2^2}\right) + u(y_1,y_2), \label{eq: 2DDiffusion} \\
&(y_1,y_2)\in [0,L_1] \times [0,L_2],~t\in[0,T], \nonumber
\end{align}
where $\alpha$ is the diffusion coefficient, $u(y_1,y_2)$ is the heat source at location $(y_1,y_2)$, and $L_1$ and $L_2$ are the domain lengths. We define the initial state field as $\xi(y_1,y_2,0)$.  We assume constant boundary conditions (no losses) by enforcing $\xi(y_1,0)=\xi(y_1,L_2)=20\mathrm{^{o}C}$ and $\xi(0,y_2)=\xi(L_1,y_2)=20\mathrm{^{o}C}$. Each nonzero source $u(y_1,y_2)$ corresponds to a control input and we have a total of $q=36$ control inputs.

We generate field data to build the DMD model by simulating the PDE system. The PDE system is simulated by using a backward spatial and temporal finite difference method  over a $71\times 71$ spatial mesh grid. A current limitation of DMD techniques is that they might not adequately capture domain boundary effects \cite{kutz2016dynamic}. To handle this issue, we focus on a sub-region inside the domain (as shown in Figure \ref{fig: 2D_diffusion}). The selected region of interest in Figure \ref{fig: 2D_diffusion} consists of a $50\times 50$ mesh that corresponds to the image pixels. The heat sources are equally spaced inside this domain. We assume that the mesh points correspond to those of the state field reported to the MPC controller  consequently, we have that the state field has a dimension of $n=2500$.  We discretize time using steps $\Delta t=1$ second and these corresponds to the sampling times of the MPC controller.  We note that there is a significant imbalance in the number of states and controls (a ratio of $36/2500=1.4\%$); consequently, we expect to see  controllability issues in this system. 

The PDE system is simulated over a time horizon over 5000 timesteps. Here, we introduce independent and random step signals (held constant over 50 timesteps) for each heating device.  These random step signals fully excite the system modes. To identify the DMD model \eqref{eq: ROM}, we choose the first $m$ timesteps of state fields and inputs training data and the remaining timesteps as validation data.  In the discussion that follows, we will explore the effect of $m$ on the accuracy of the DMD model and on the performance of the DMD-controller.

\subsection{DMD Model Accuracy}
\begin{figure}[!htp]
\begin{center}
	\includegraphics[width=3in]{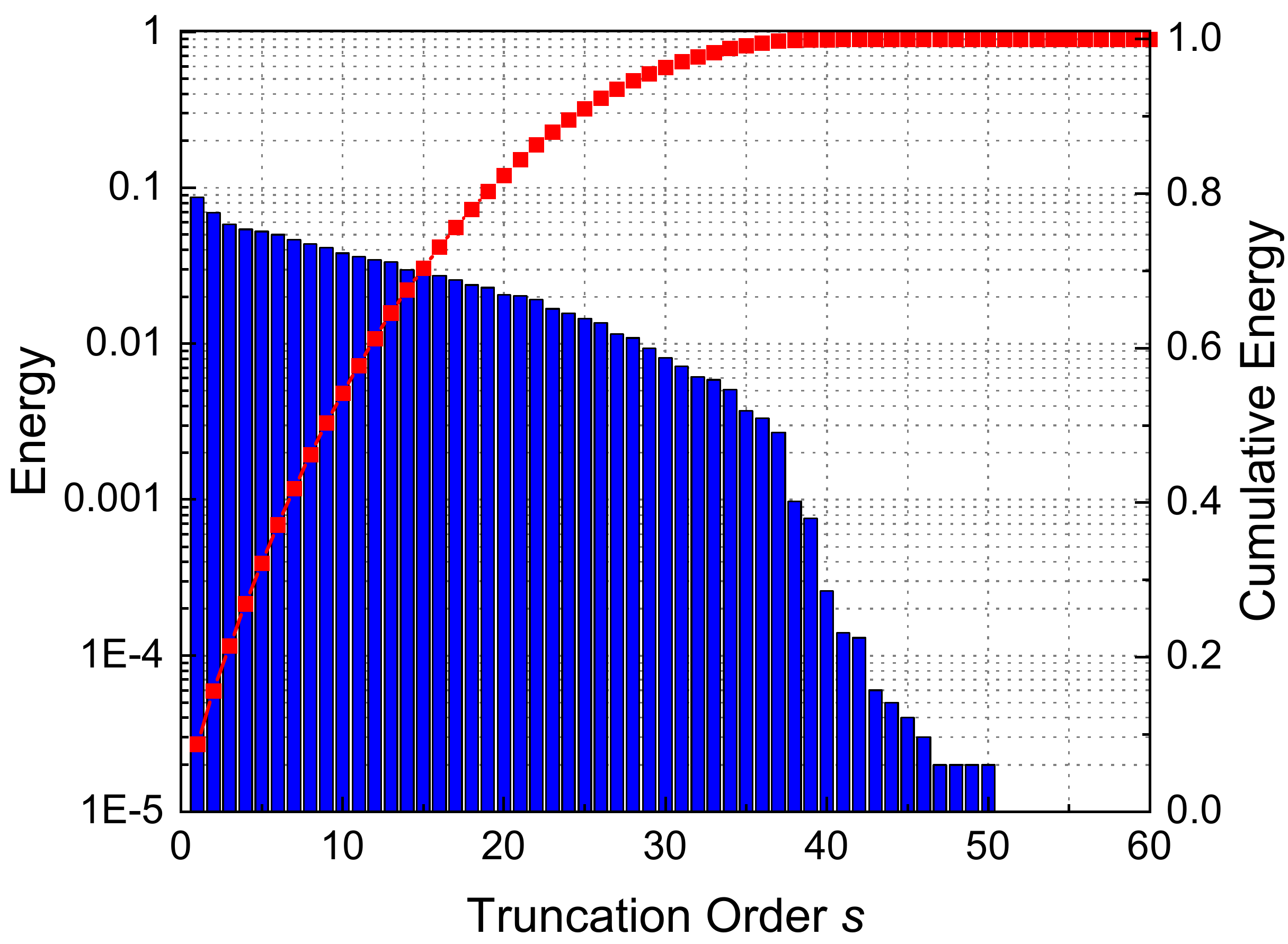}\qquad
	\includegraphics[width=3in]{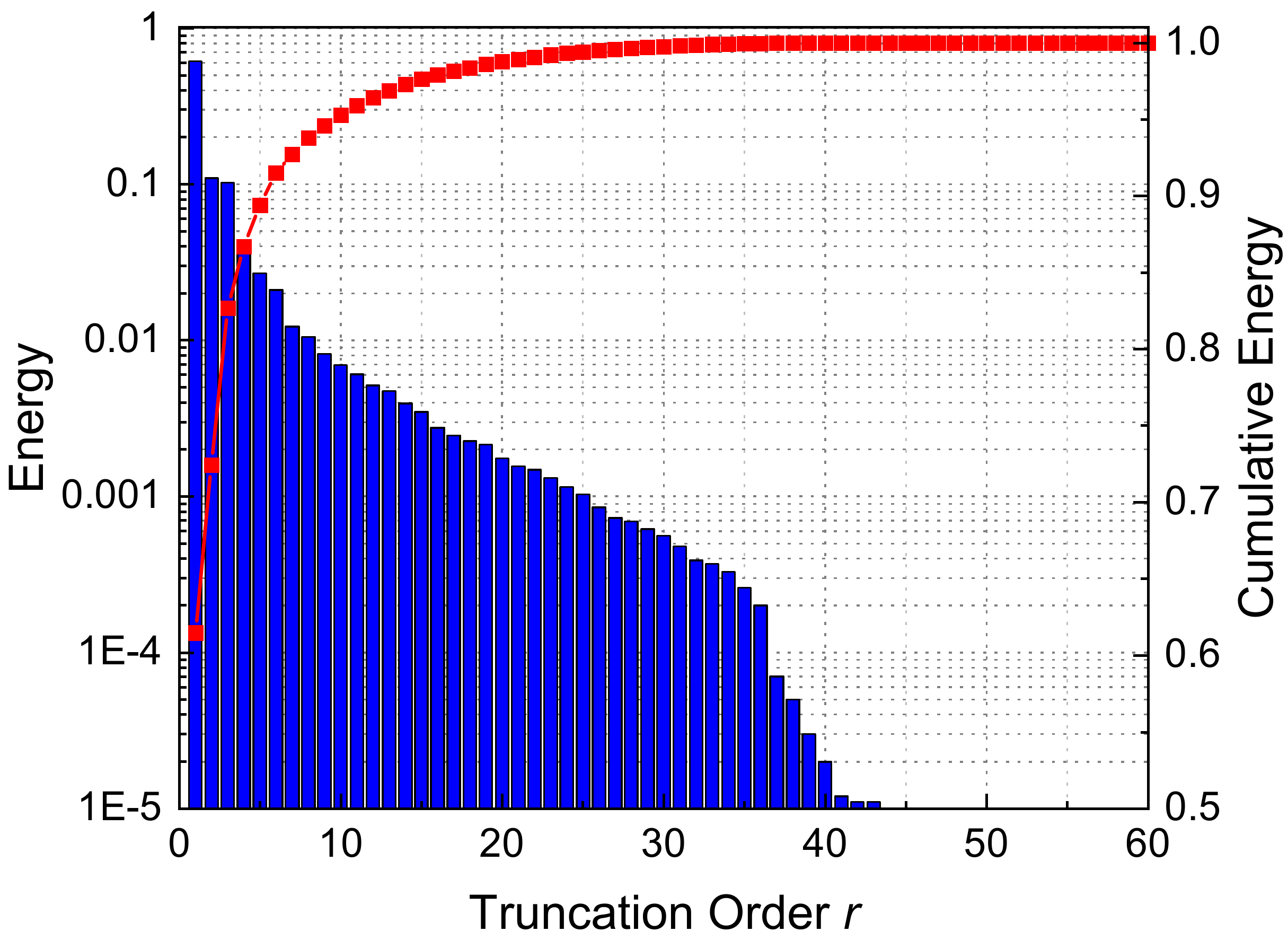}\caption{Energy for different SVD  truncation orders in \eqref{eq: SVD_Omega} (left) and in \eqref{eq: SVD_Y} (right).} \label{plot: Truncation}
\end{center}
\end{figure}

The DMD bases are obtained from \eqref{eq: SVD_Omega} and \eqref{eq: SVD_Y}. The singular values associated with such bases quantify the importance of each basis vector in capturing the information (energy) of the  system. We define a cumulative energy criterion to determine appropriate model reduction orders $s$ and $r$:
\begin{equation}
p_{v}=\frac{\displaystyle\sum_{i=1}^{v}\sigma_{i}^2}{\displaystyle\sum_{i=1}^{k}\sigma_{i}^2},
\end{equation}
where $k=\min\{n+q,m\}$ and $v=s$ for SVD \eqref{eq: SVD_Omega} and $k=\min\{n,m\}$  and $v=r$ for SVD \eqref{eq: SVD_Y}. The $i$-th singular values for both SVDs are denoted by $\sigma_i$. We use the energy $p_v$ to determine suitable model truncation orders. The orders should be chosen such that $p_v$ captures a  sufficiently large portion of the energy. A rule-of-thumb threshold for determining the orders is $p_v=0.99$. In the context of DMD, the accuracy of the first SVD impacts that of the second SVD and thus in this example, we tend to choose a higher threshold for the first SVD. In Figure \ref{plot: Truncation}, we present values of $p_v$ against different SVD truncation orders. We can see that a model order $s=50$ accounts for $99.82\%$ of the variation in the data matrix $\mathbf{\Omega}$ and $r=40$ accounts for $99.70\%$ of the energy in $\mathbf{X}$. The reduced model \eqref{eq: ROM} with such a selection of parameters is a non-square state-space model with $r=40$ states and $q=36$ inputs.  We use this model order as default; in the discussion that follows, we also explore the behavior of DMD-MPC with different model orders. We recall that the low-dimensional state of the reduced model $\tilde{\mathbf{x}}_k$ is projected to the original state space by using the DMD bases. 

\begin{figure}[!htp]
 \begin{center}
	\includegraphics[width=7in]{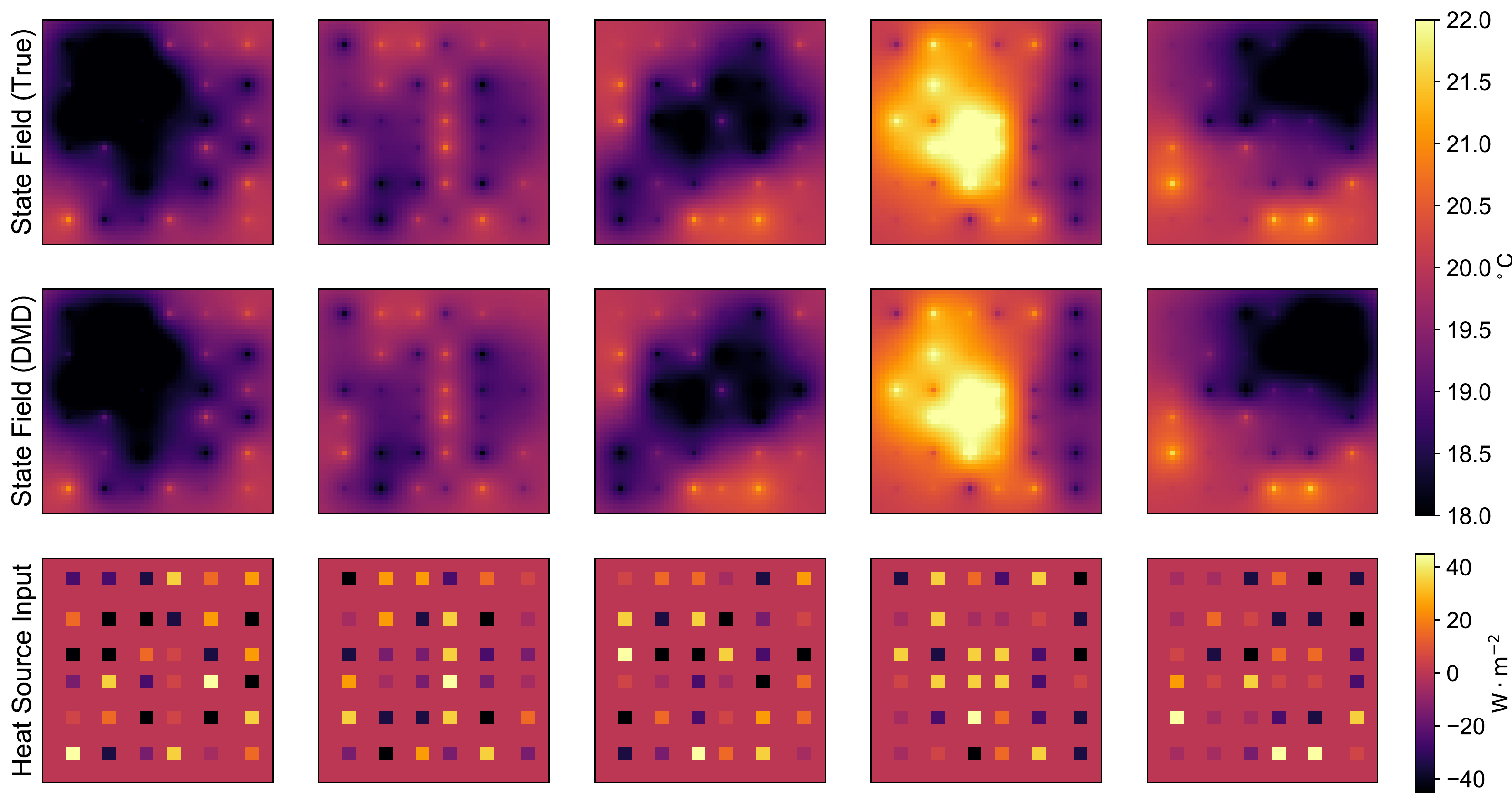}\caption{ True state field (top), predicted state field with  DMD model (middle), and corresponding heat inputs (bottom).}\label{fig: DMDPrediction}
 \end{center}
\end{figure}

%\newpage

Figure \ref{fig: DMDPrediction} compares the predicted DMD field \eqref{eq: ROM} against the actual state field at five different times. Here, we use $m=3000$ image snapshots to build a DMD model of order 40. We can see that the predictions of the DMD model are accurate; specifically, we found that the maximum absolute error (over all spatial locations) was $\mathcal{O}(10^{-3})$. This result indicates that a model of order 40 can capture the behavior of the full-order system (order 2500). This represents a reduction in the state space of 98.5\%, thus enabling the use of MPC.  The model accuracy will be key in enabling the MPC controller to enforce system constraints. We highlight that recent work has established the equivalence between DMD and subspace system identification \cite{shin2020unifying}; we thus expect that the  performance obtained with such models would be comparable. 

\textbf{Remark 3}. In general, identifying a low-order model from high-dimensional data requires significantly less data than identifying the full-order model (similar in spirit to subspace identification techniques widely used in industrial applications of MPC).  Specifically, for a linear dynamical system with low rank and, in the absence of noise, the dominant spatial and temporal modes can be identified with few samples if the experimental input excites the system sufficiently. However, when using DMD to approximate nonlinear systems, exciting more modes can assist to discover all potential significant modes that are important for improving prediction accuracy of developed low-order model. Thus, a relative large number of data samples is preferred for selecting a suitable order. Moreover, the presence of measurement noise can decrease the signal-to-noise ratio in the data, which also demands more data samples to mitigate the effect of noise and discover the hidden modes.  Finally, we observe that the proposed approach is to be implemented in real-time, and thus data can be accumulated over time to progressively refine the model.
%\begin{figure}[tbh]
%	\begin{center}
%		\includegraphics[width=3in]{TrueAverage.eps}
%		\includegraphics[width=3in]{ErrorProfile.eps}
%		\caption{Left: Averaged absolute profile of the true state values in the test data; Right: Averaged absolute error profile between DMDc reconstruction and true state field.}\label{fig: ErrorProfile}
%	\end{center}
%\end{figure}

\subsection{Closed-Loop DMD-MPC Behavior}\label{sec:MPCDiffusion}

The MPC formulation uses a DMD model of order $r=40$ and has $q=36$ control inputs. The formulation captures lower and upper bounds for the states and controls (the sets $\mathcal{X}$ and $\mathcal{U}$ are boxes). The prediction horizon of the MPC controller is set to $N=10$; consequently, the MPC formulation contains a total of $10\times (40)+(10-1)\times 36=724$ variables.  An MPC controller that operates over the full state-space would contain 25,324 variables. Importantly, such a problem would contain dense blocks in the dynamic constraints that will dramatically affect the computational tractability. 

\begin{figure}[tbh]
	\begin{center}
		\includegraphics[width=0.7\textwidth]{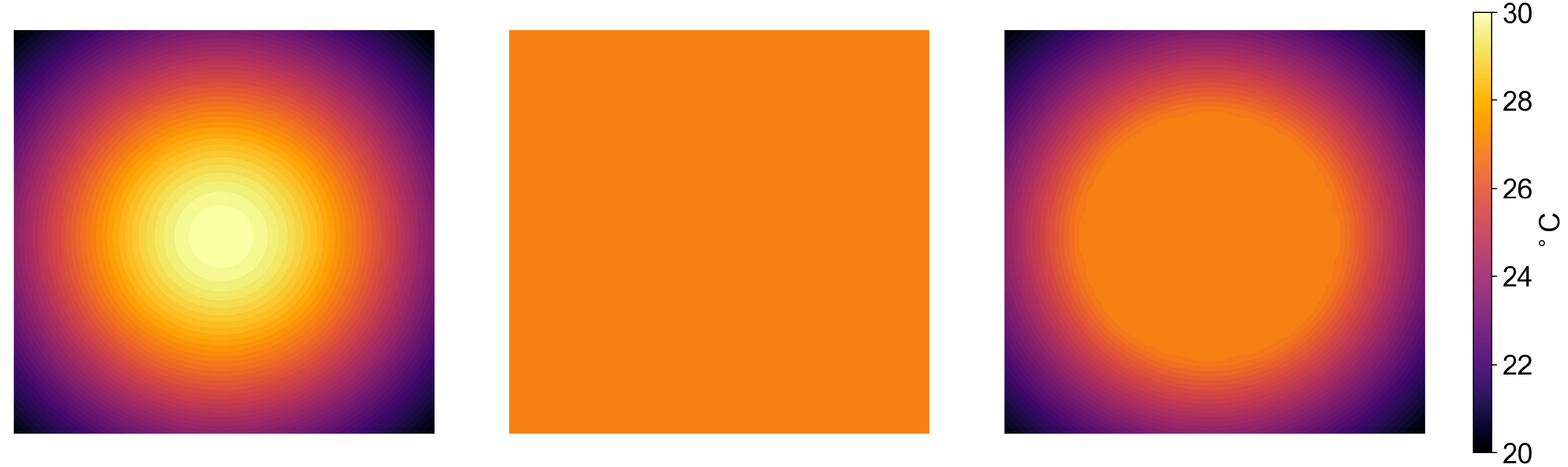}
		\caption{Reference fields used to test performance of controllers. Gaussian reference field (left), constant reference field (middle), and sliced Gaussian field (right).}\label{fig: References}
	\end{center}
\end{figure}

\begin{figure}[tbh]
	\begin{center}
		\includegraphics[width=0.8\textwidth]{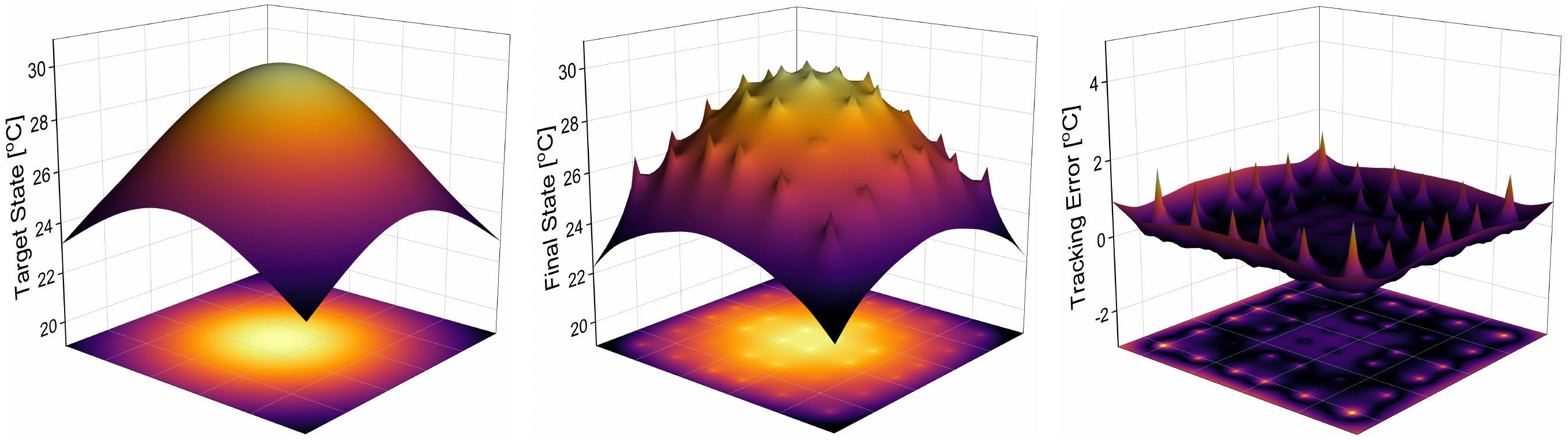}
		\caption{Reference field (left), final state field for closed-loop system under DMD-MPC (middle),} and final tracking error achieved by DMD-MPC (right).\label{fig: StateTarget}
	\end{center}
\end{figure}

Figure \ref{fig: References} shows three reference fields that will be used to test the performance of the DMD-MPC controller. We assume that the system is initially at steady-state. We first set the target field to be the Gaussian field shown on the left plot in Figure \ref{fig: StateTarget} and we run the closed-loop system over 30 timesteps. In the middle plot of Figure \ref{fig: StateTarget}, we show the final temperature field achieved by DMD-MPC and in the right plot, we show the error field (difference between the target and DMD-MPC fields). We can see that the controller is able to capture the global (coarse) features of the target field but has localized (fine) errors at the actuator locations. The local spikes indicate that the controller sacrifices some local error in order to reduce global error (i.e., in order to drive the entire thermal field up). We highlight that the fact that the controller cannot perfectly track the target is not due to the model reduction procedure but it is due to the fact that the original system has limited controllability. To validate this observation, we performed a simulation in which the controller uses the model of the real system and we found similar closed-loop performance. This highlights that the availability of image data only benefits the ability to control the system up to a certain point (limited by its inherent controllability). 

\begin{figure}[!htp]
	\begin{center}
		\includegraphics[width=0.8\textwidth]{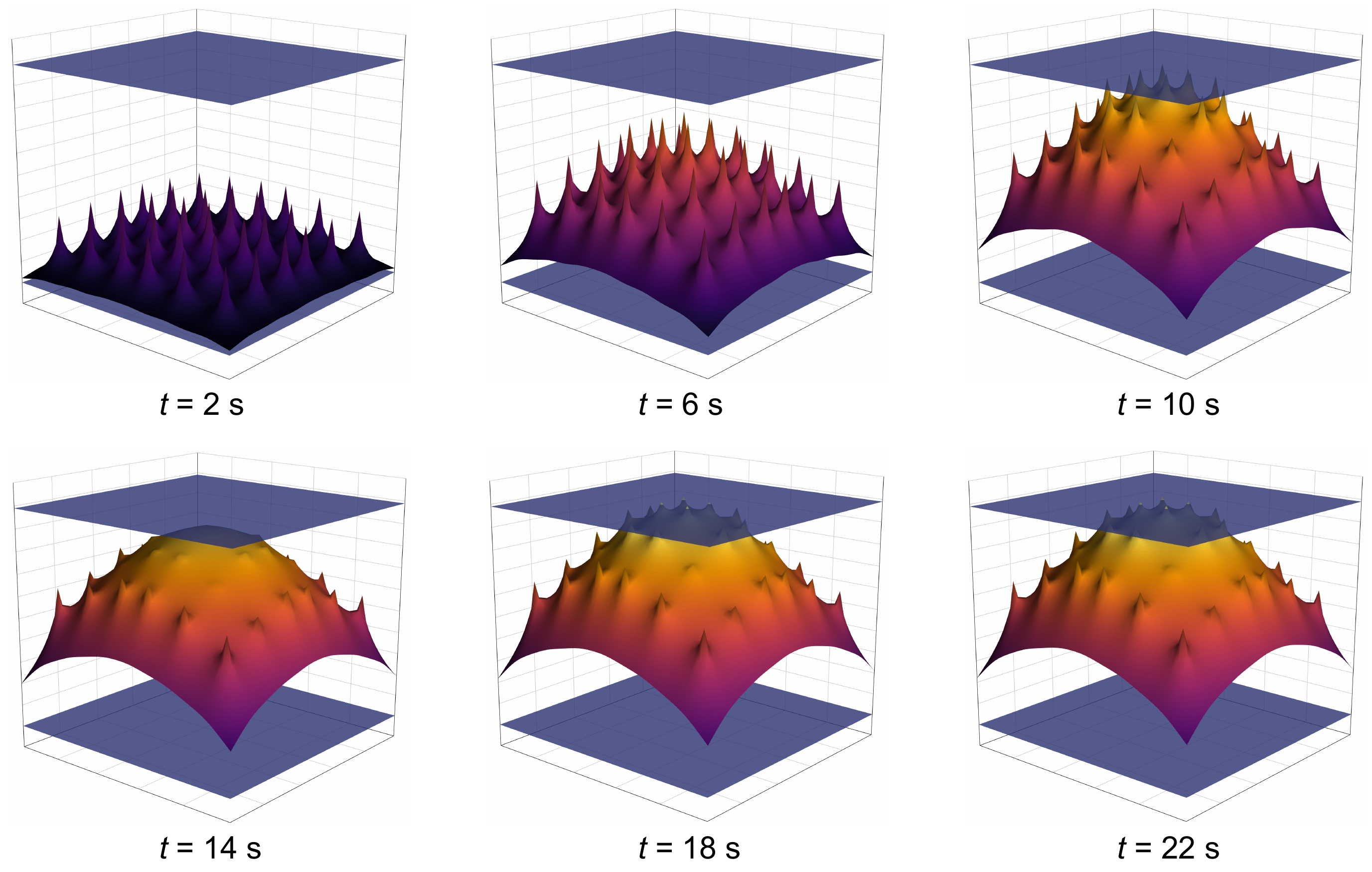}
		\caption{State evolution of diffusion system under DMD-MPC showing satisfaction of constraints.}\label{fig: StateEvolution}
	\end{center}
\end{figure}

\begin{figure}[!htp]
	\begin{center}
		\includegraphics[width=\textwidth]{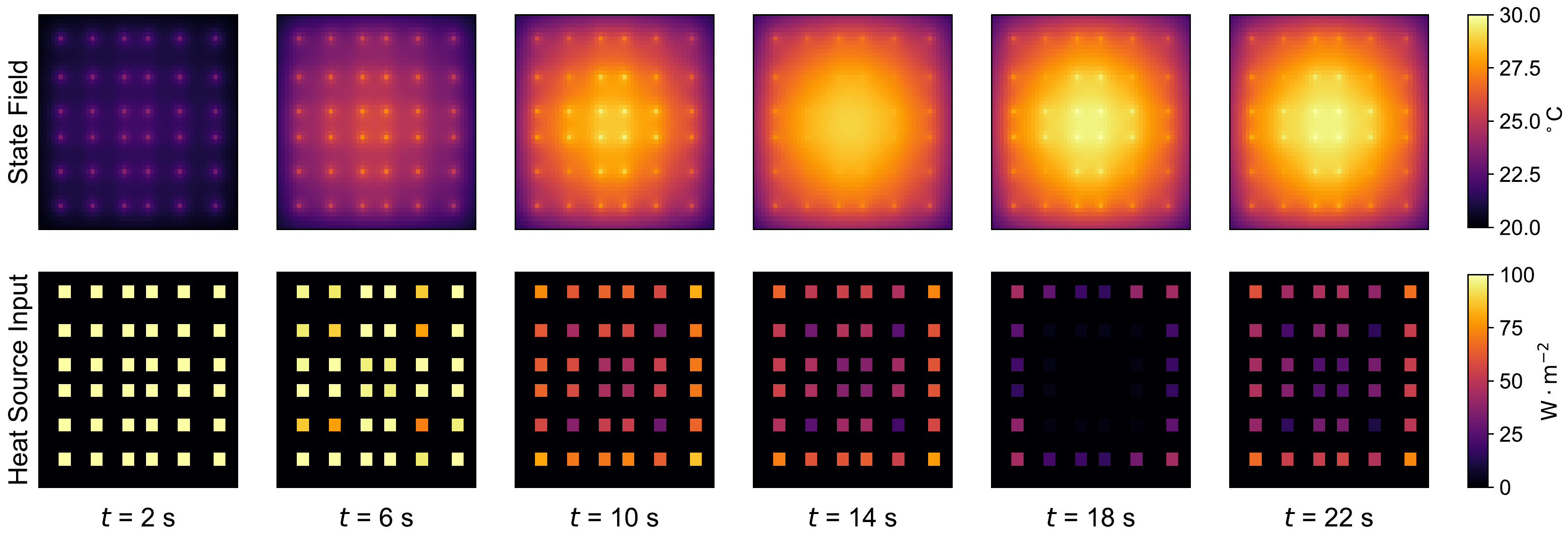}
		\caption{Input and state evolution of diffusion system under DMD-MPC.}\label{fig: InputEvolution}
	\end{center}
\end{figure}

\begin{figure}[!htp]
	\begin{center}
		\includegraphics[width=0.8\textwidth]{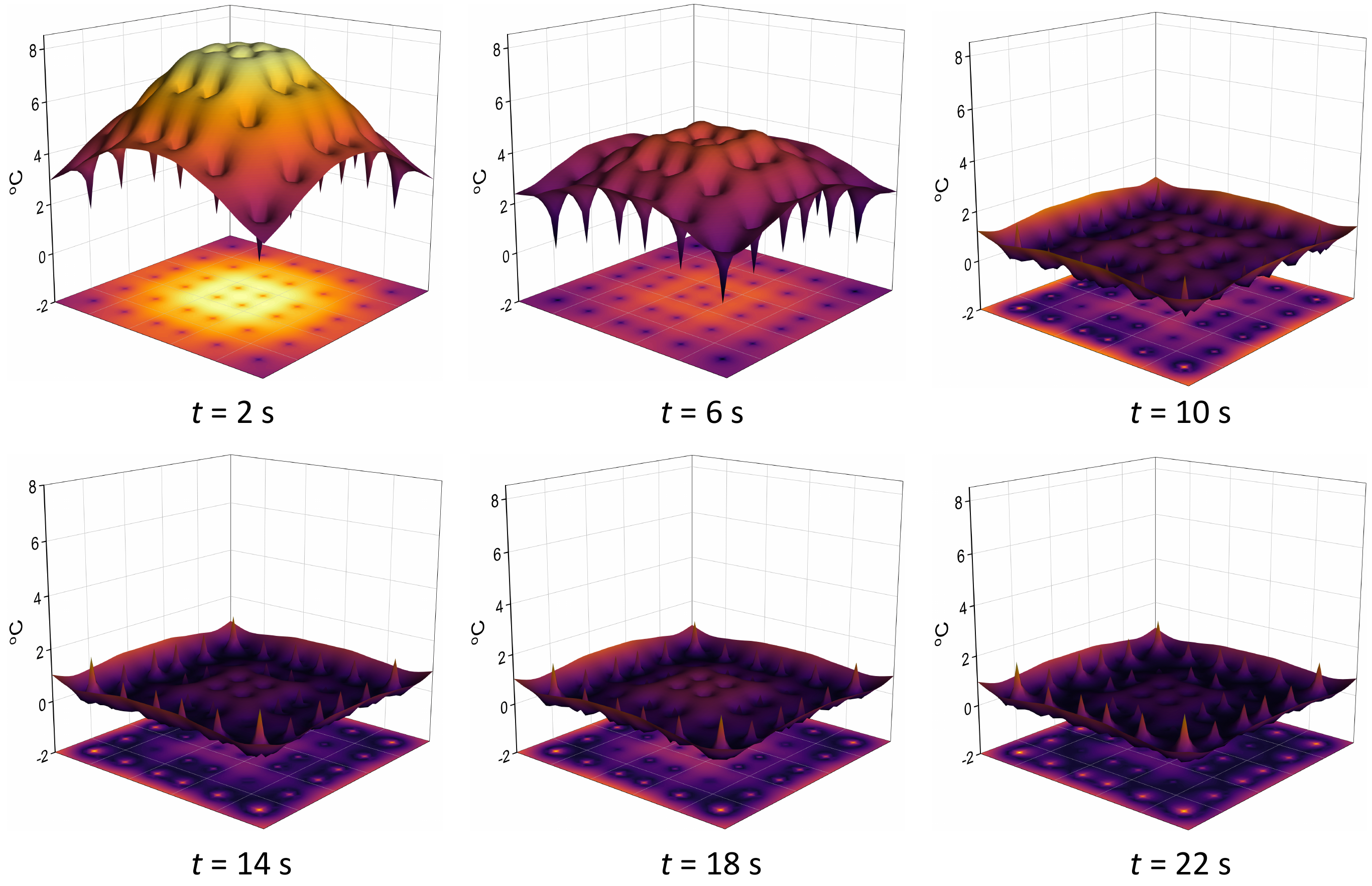}
		\caption{Tracking error field of diffusion system under DMD-MPC.}\label{fig: ResidualEvolution}
	\end{center}
\end{figure}

Figure \ref{fig: StateEvolution} shows the time evolution of the thermal fields. Here, we visualize the fields as 3D fields (intensity is the third dimension) in order to verify the satisfaction of state constraints (bottom and top planes). We found that the DMD-MPC controller satisfies the constraints almost all the time, with only a few locations that violate the upper bound at time $t=22s$. This reinforces the observation that the DMD model is accurate. Figure \ref{fig: InputEvolution}  shows the images of the state field and the corresponding heating device profiles. The actuators operate at the maximum allowable value in the first few time steps to drive up the state profile rapidly and then settle down to the steady-state. The evaluation of the tracking error over time is shown in Figure \ref{fig: ResidualEvolution}. Here, we can see that the errors decrease quickly and that the final error consists of high-frequency spikes around actuator locations and at the boundaries. 
\\

\begin{figure}[!htp]
	\begin{center}
		\includegraphics[width=0.7\textwidth]{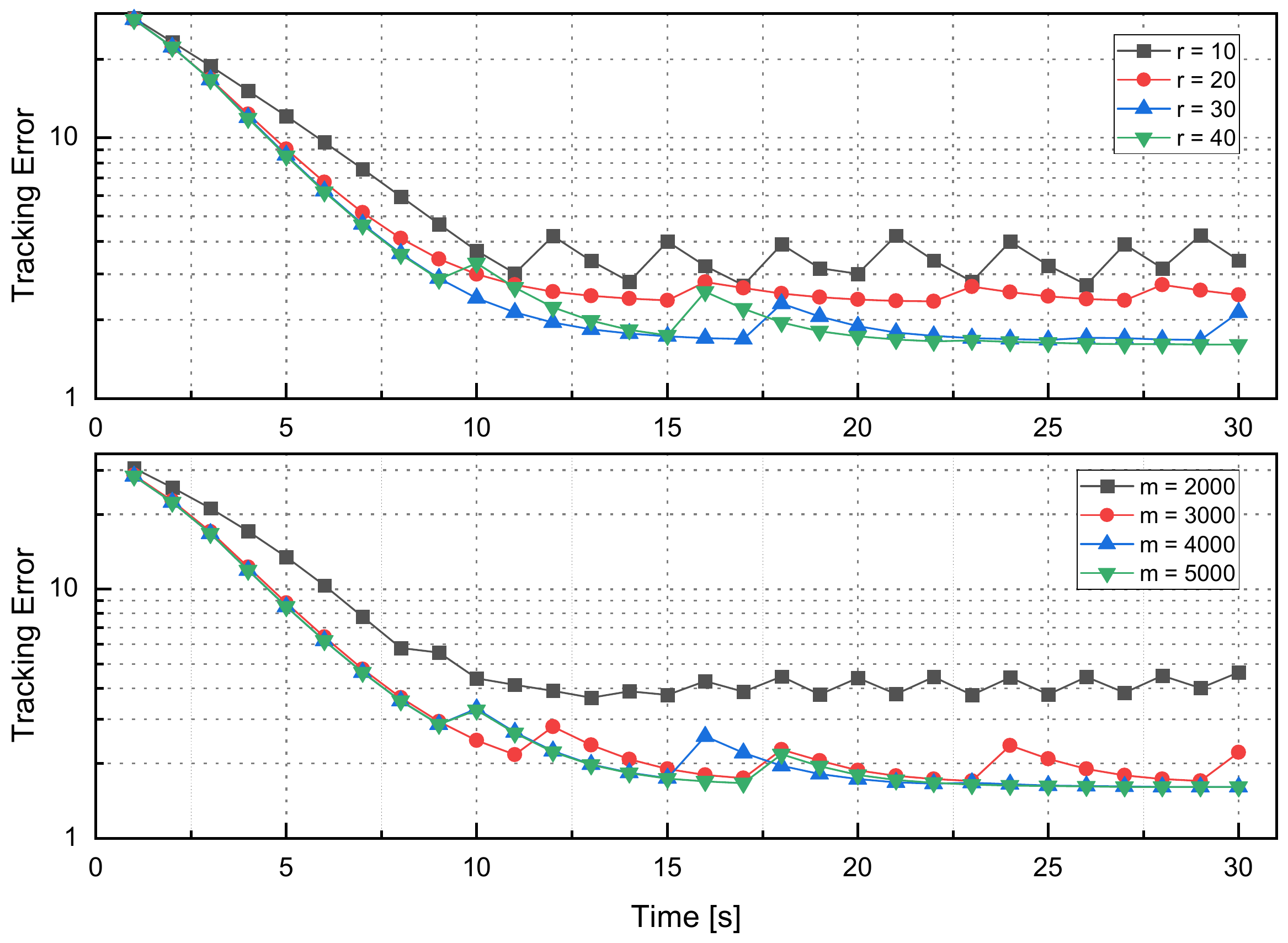}
		\caption{Impact of model order and dataset size on tracking performance of DMD-MPC. }\label{fig: compare}
	\end{center}
\end{figure}

We investigated the impacts of data availability $m$ and model orders $r,s$ on the performance of DMD-MPC. Figure \ref{fig: compare} shows the tracking error under different settings, where the error is defined as the 2-norm of the difference between actual state and the reference: $\|\mathbf{e}_{k}\|_2=\|\mathbf{x}^{*}-\mathbf{x}_{k}\|_2$. We can see that, as the model order increases, the tracking error decreases. Moreover, we see that orders 30 and 40 show comparable performance. The bottom plot of Figure \ref{fig: compare} shows that data availability has a strong impact on tracking performance. Intuitively, a larger dataset enables the discovery of spatiotemporal modes of the system. We also note that we require at least 2000 image snapshots to obtain adequate accuracy. 
\\

\begin{figure}[!htp]
	\begin{center}
		\includegraphics[width=\textwidth]{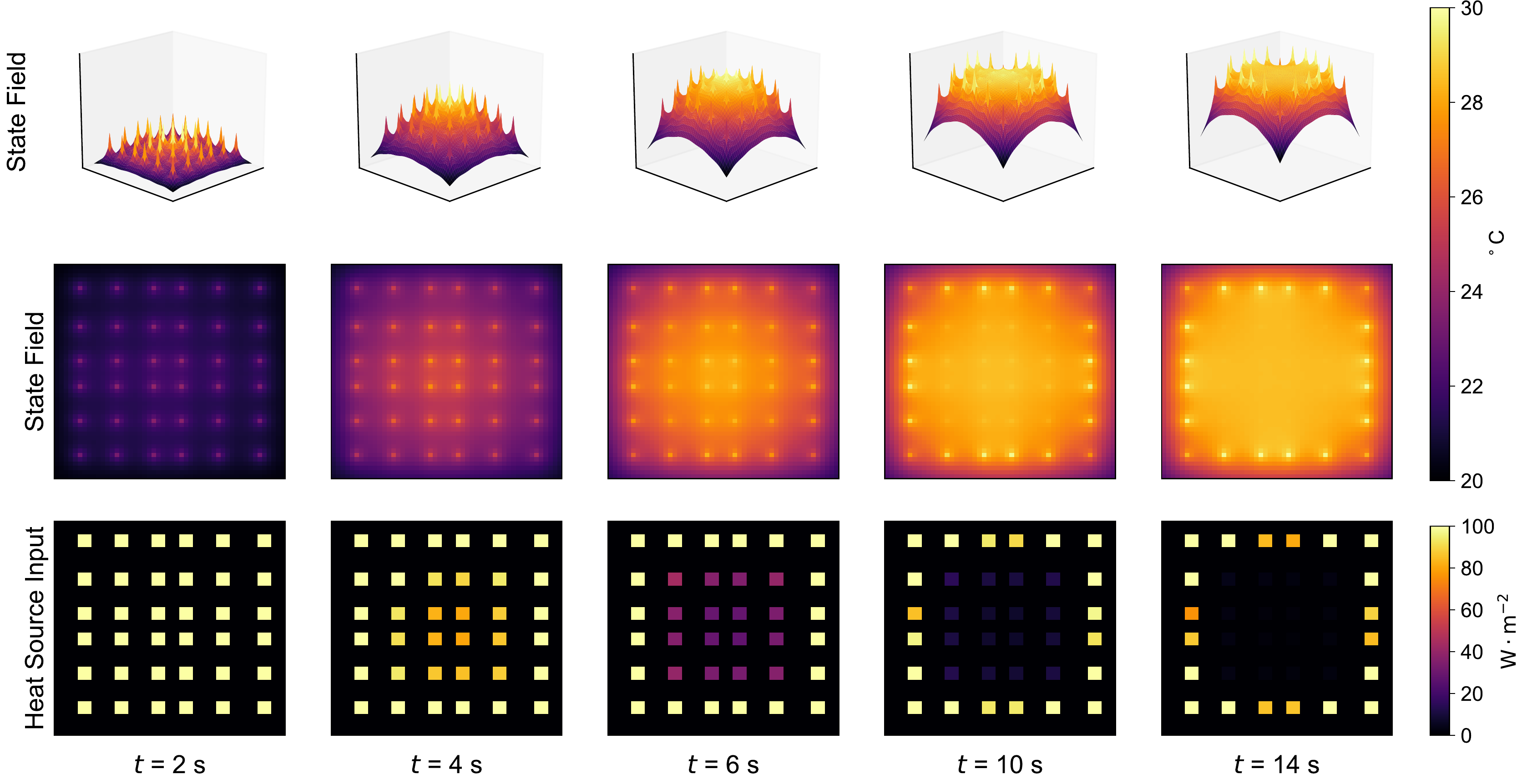}
		\caption{Input and state evolution under DMD-MPC with constant reference field. }\label{fig: ConstantSetpoint}
	\end{center}
\end{figure}

We also investigated the performance of the closed-loop system under a constant reference target $\mathbf{x}^{*}=28\mathrm{^o C}$ (shown in the middle plot of Figure \ref{fig: References}). The evolution of state and input trajectories for this reference is shown in Figure \ref{fig: ConstantSetpoint}. We can see that the controller is able to track the target with satisfactory performance (except at the boundaries). 

\begin{figure}[!htp]
	\begin{center}
		\includegraphics[width=0.6\textwidth]{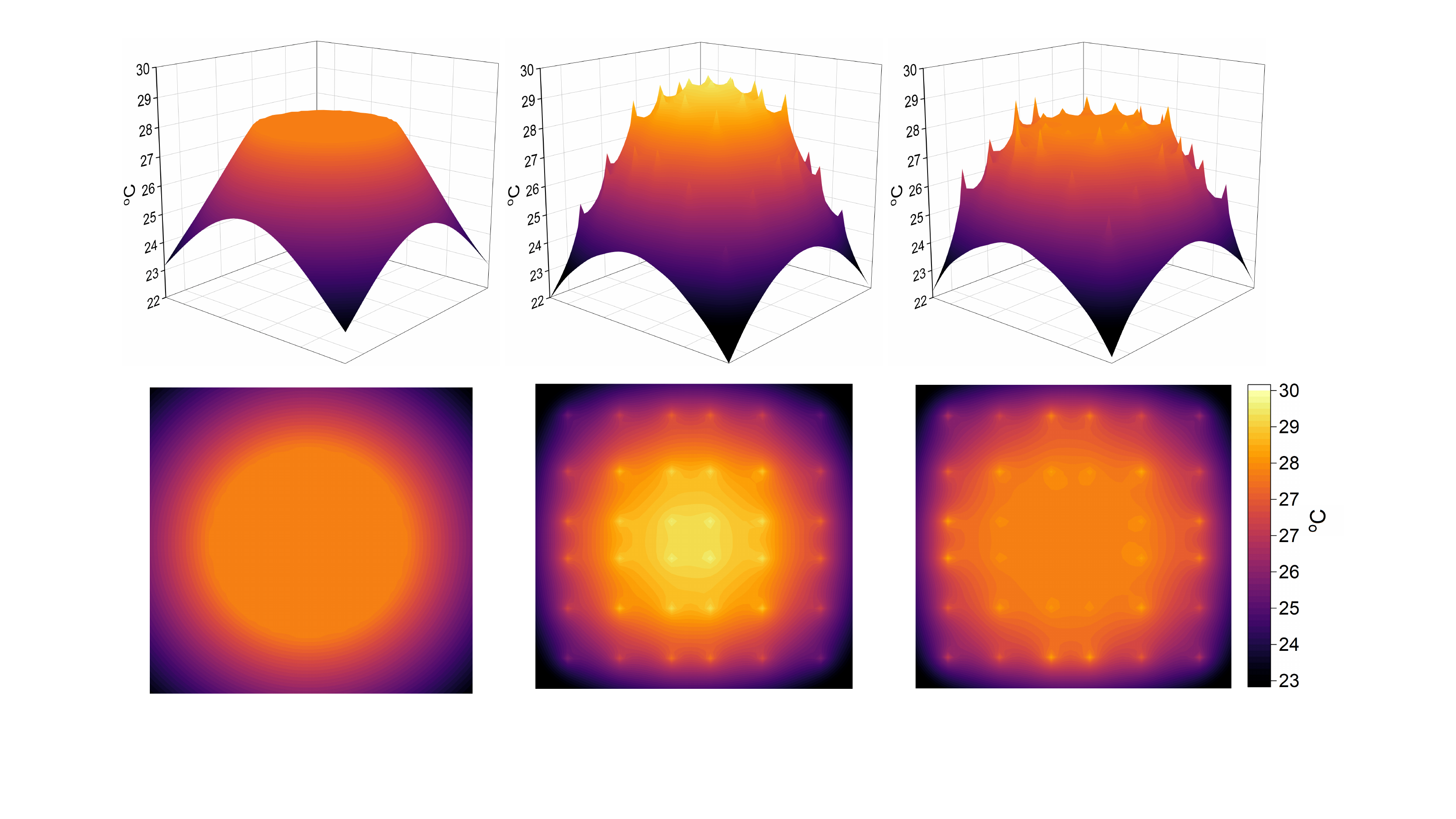}
		\caption{Target field for DMD-MPC and standard MPC (left), final state field achieved with  standard MPC (middle), and final state field achieved with DMD-MPC (right).}\label{fig: TraditionalMPC}
	\end{center}
\end{figure}

\begin{figure}[!htp]
	\begin{center}
		\includegraphics[width=0.9\textwidth]{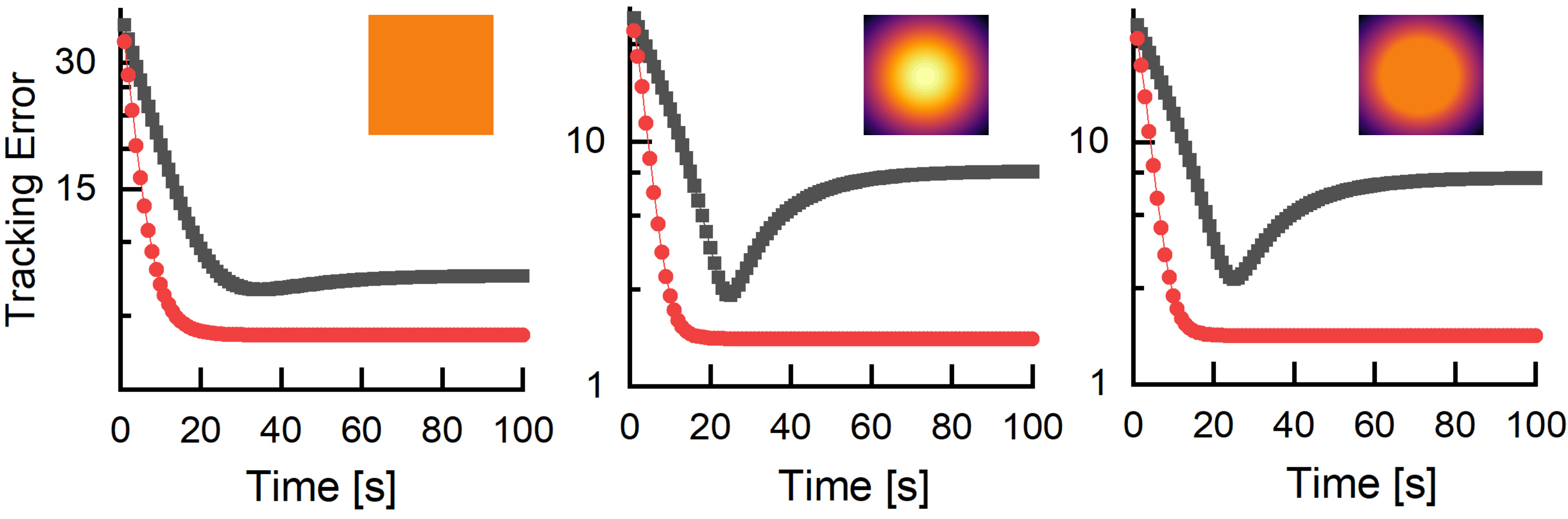}
		\caption{Tracking error for different reference fields under DMD-MPC (red) and standard MPC (black).} \label{fig: TraditionalVSDMD_MPC}
	\end{center}
\end{figure}

\subsection{Comparison with Standard MPC}

Standard MPC aims to control the state field by controlling the temperature at a fixed set of locations (proxy points). Importantly, such locations are often a poor low-dimensional representation of the system. Specifically, such locations might not capture the global effects properly. In contrast, the DMD modes capture the global effects of the field. For developing the standard MPC controller,$  $ we assume that there exists a temperature sensor at the location of each heating device. We use DMD to identify a state-space model based on data collected at such locations. In this case, the DMD model has a dimension of $n=36$ (corresponding to the number of devices). We highlight that this DMD model is a {\em multivariable} system model that captures interactions between the temperatures at different locations. We will see, however, that these interactions are not sufficient to capture the complexity of the high-dimensional state field. We also highlight that the dimension of the model used by the standard MPC controller is comparable to that used by DMD-MPC. 
\\

We compare the tracking performance of standard MPC with that of the DMD-MPC in Figure \ref{fig: TraditionalMPC}. Here, we use the sliced Gaussian reference field shown in Figure \ref{fig: References}. The plots in the middle of Figure \ref{fig: TraditionalMPC} show the final state field achieved by standard MPC. From the middle plots of Figure \ref{fig: TraditionalMPC}, we observe that standard MPC is not able to track the flat part of the target profile, as indicated by the large peaks in the center of the domain. This indicates that the model cannot capture the global features of the state field.  In contrast, the model of DMD-MPC captures the global features of the field (contained in the modes) and this gives much better performance (as shown in the right plots of Figure \ref{fig: TraditionalMPC}). 
\\

In Figure \ref{fig: TraditionalVSDMD_MPC}, we compare the tracking error of the MPC controllers under different reference fields.  We observe that for all references,  DMD-MPC consistently achieves a smaller tracking error. Moreover, we see that the final tracking error of DMD-MPC is comparable in all cases. We also see that, for a constant field,  the performance of the controllers is comparable. This makes sense because a constant reference field is much easier to reach. However, the discrepancy in controller performance is quite strong for the more complex reference fields. These results illustrate the value embedded in image data. Specifically, using MPC controllers that do not have access to such data can miss significant information of the state field and this can limit their ability to manipulate it. 

%%%%%%%%%%%%%%%%%%%%%%%%%%%%%%%%%%%%%%%%%%

\section{Conclusions and Future Work} \label{sec: Summary}

We have presented a data-driven MPC framework that uses low-order models (constructed from spatial image data) to control high-dimensional state fields. Here, we show that DMD facilitates the exploitation of image data to construct accurate low-order models. We demonstrate the benefits of the proposed DMD-MPC framework by using a 2D thermal diffusion system. We show that a DMD model of order 40 is sufficient to accurately predict the performance of a system of dimension 2500. We also show that the DMD-MPC controller provides satisfactory tracking and constraint satisfaction performance. We compare the performance of DMD-MPC against that of a standard MPC controller that uses data from a finite set of locations. We find that such a controller is incapable of properly capturing the features of the high-dimensional field (thus resulting in poor tracking performance). The proposed work can be extended in multiple directions; specifically, we will aim to investigate the use of more advanced MPC formulations (economic MPC) and we will aim to construct robust MPC formulations that capture the model error explicitly. We will also seek to construct low-order models that provide certifiable controllability guarantees and that can fuse data-driven and physics-based components. 

%%%%%%%%%%%%%%%%%%%%%%%%%%%%%%%%%%%%%%%%%%

\section*{Acknowledgments}
We acknowledge financial support from the University of Wisconsin-Madison. 

%%%%%%%%%%%%%%%%%%%%%%%%%%%%%%%%%%%%%%%%%%

\bibliography{vzavala}

\end{document}